\documentclass[showpacs,floatfix,twocolumn]{revtex4-1}
\usepackage{graphicx,psfrag,amsmath,amssymb,amsfonts,latexsym,color,epsf,dcolumn,graphpap}
\usepackage{subfig}
\usepackage{float}
\usepackage{lipsum}
\usepackage{enumerate}

\usepackage{tikz-cd} 
\usepackage{tikz}
\usetikzlibrary{shapes.geometric, arrows}

\definecolor{red}{rgb}{1,0,0}
\definecolor{blue}{rgb}{0,0,1}
\definecolor{skyblue}{rgb}{0,0,.5}
\definecolor{green}{rgb}{0,1,0}
\definecolor{orange}{cmyk}{0,.4,1,0}

\begin{document}
\title{Shortcut to adiabaticity in a cavity with a moving mirror}

\author{Nicol\'as F.~Del Grosso$^1$ }
\author{Fernando C. Lombardo$^1$}
\author{Francisco D. Mazzitelli$^2$}
\author{Paula I.~Villar$^1$ }
\affiliation{$^1$ Departamento de F\'\i sica {\it Juan Jos\'e
 Giambiagi}, FCEyN UBA and IFIBA CONICET-UBA, Facultad de Ciencias Exactas y Naturales,
 Ciudad Universitaria, Pabell\' on I, 1428 Buenos Aires, Argentina }
\affiliation{$^2$ Centro At\'omico Bariloche and Instituto Balseiro,
Comisi\'on Nacional de Energ\'\i a At\'omica, 
R8402AGP Bariloche, Argentina}

\begin{abstract}
\noindent Shortcuts to adiabaticity constitute a powerful alternative that speed up time-evolution while mimicking adiabatic dynamics. In this paper we describe how to implement shortcuts to adiabaticity for the case of a massless scalar field inside a cavity with a moving wall, in 1 + 1 dimensions. The approach is based on the known solution to the problem that exploits the conformal symmetry, and the shortcuts take place whenever there is no dynamical Casimir effect. We obtain a fundamental limit for the efficiency of an Otto cycle with the quantum field as a working system, that depends on the maximum velocity that the mirror can attain. We describe possible experimental realisations of the shortcuts using superconducting circuits. 
\end{abstract}
\date{today}
\maketitle
\noindent 

\maketitle
\section{Introduction}\label{sec:intro}

In recent times, as new technologies allow us to manipulate smaller
and smaller systems, such as trapped ions, nano resonators and electronic
circuits \cite{ion,circuitqed,nanoresonator}, a natural question has emerged about whether it is possible
to use them to produce machines and what their properties would be.
The novelty comes from the fact that these small systems can exhibit
quantum properties that could potentially be exploited to get an advantage
over classical machines or present new obstacles to the operation.
These questions constitute the back bone of a new area of physics
that has come to be called quantum thermodynamics. 
In some cases of
discrete stroke quantum machines, such as a quantum harmonic oscillator
or a quantum field undergoing an Otto cycle, it has been shown that the
efficiency of the resulting machine is maximum for an adiabatic (i.e.
infinitely slow) driving \cite{fields1,Otto_nos}. The problem is that under these conditions
the power of the machine vanishes and so it becomes necessary to understand
how to operate these machines in a finite time. 
However this leads to a friction work on each stroke $\langle w_{\text{fric}}\rangle=\langle w\rangle-\langle w_{\text{ad}}\rangle$, defined as the difference between the actual work and the adiabatic work, that is always non-negative \cite{friction}. This has the effect of reducing the efficiency of a heat engine. For example, for a quantum Otto cycle comprised of four stokes (A, cooling at constant volume; B, adiabatic expansion; C, heating at constant volume; D, adiabatic compression) the efficiency in finite time is given by 
\begin{equation}
    \eta=\frac{W}{Q}=\frac{W_{\text{ad}}-\langle w_{\text{fric}}\rangle_{AB}-\langle w_{\text{fric}}\rangle_{CD}}{Q_{\text{ad}}-\langle w_{\text{fric}}\rangle_{AB}}\leq\frac{W_{\text{ad}}}{Q_{\text{ad}}}=\eta_{\text{ad}},
\end{equation}
where $W_{\text{ad}}=\langle w_{\text{ad}}\rangle_{AB}+\langle w_{\text{ad}}\rangle_{CD}$, and is always lower or equal to the adiabatic efficiency. Therefore, it is paramount to understand if it is possible to implement an adiabatic evolution in finite time.

Although in most cases
the finite time operation causes the emergence of coherences in the
state of the system that result in an efficiency loss, in many cases
it is possible to implement protocols, named shortcuts to adiabaticity
(STA), that evolve the initial state into the final state that would
have been obtained with an adiabatic evolution,  but in a finite time \cite{berry,STA1,STA2,STA3}. These protocols typically require a full control of the quantum system
and end up being extremely challenging from an experimental stand
point.

In previous works, STA have been considered 
from a theoretical and/or an experimental point of view for different physical systems: trapped ions \cite{Palmero}, cold atoms \cite{Torrontegui}, ultracold Fermi gases \cite{Dowdall}, Bose-Einstein condensates in atom chips \cite{Amri}, 
spin systems \cite{baris}, etc.  STA have been also proposed to relieve the trade-off of efficiency and power \cite{engine1,engine2,engine3}, both in single-particle quantum heat engines (QHE) \cite{delcampo1},  and in many-particle QHE \cite{delcampo2}. 
There was even an experiment with a unitary Fermi gas that implemented last systems \cite{delcampo3}. On the other hand, authors in Ref. \cite{delcampo4} have considered many particle theories for QHE in the adiabatic case. STA's have also been obtained for relativistic quantum systems evolving under Dirac dynamics \cite{Deffner1,Deffner2}.

In this paper we explore the possibility of applying STA in quantum field theory. In particular, we will consider a scalar quantum field in a one-dimensional
cavity with a moving wall, whose state undergoes a unitary evolution.
We will show that given a wall trajectory $L_{\rm ref}(t)$ we can find a shortcut
to adiabaticity given by an effective trajectory $L_{\text{eff}}(t)$
that, when implemented in finite time, results in the same state as
if the original had been performed adiabatically. The STA occurs whenever there is no dynamical Casimir effect (DCE). 
As we will see, this protocol has the
advantage that it can be easily implemented experimentally using superconducting circuits,  since it
does not require additional exotic potentials. Moreover, the effective trajectory can be computed from
the original quite simply,  paving the way for more efficient quantum
field thermal machines.

\section{Adiabatic shortcuts for dynamical Casimir effect}

There are several ways of introducing adiabatic shortcuts
in quantum systems. A natural approach is to impose the instantaneous eigenstates of a time dependent 
Hamiltonian to be solutions of a time-dependent Schr\"odinger equation of a different Hamiltonian,
which is the original one plus a ``counter diabatic" Hamiltonian \cite{STA2}. This procedure leads in general 
to an effective non-local Hamiltonian which is difficult to implement experimentally. For some systems (such 
as a harmonic oscillator  with a time-dependent frequency $\omega(t)$), an additional unitary transformation 
simplifies the effective Hamiltonian, without modifying the initial and final states \cite{STA5}. Also for these systems, an 
alternative approach \cite{calzetta} is based on the following observation: the adiabatic WKB solution for the position operator 
$\hat q(t)$ can be written in terms of annihilation and creation operators as
\begin{equation}
\hat q(t)= \hat a \frac{e^{- i \int^t\omega_{\rm ref}(t')dt'}}{\sqrt{2\omega_{\rm ref}(t)}}+\hat a^\dagger \frac{e^{ i \int^t\omega_{\rm ref}(t')dt'}}{\sqrt{2\omega_{\rm ref}(t)}}
\end{equation}
(we stress that along this paper we will be working with natural units where $c=\hbar=k_B=1$).
This is an approximate solution for the oscillator
with a reference frequency $\omega_{\rm ref}(t)$, valid if it is slowly varying,  but an
exact solution of a system with an effective frequency \cite{calzetta}
\begin{equation}
\omega_{\rm eff}^2(t)=\omega_{\rm ref}^2+\frac{1}{2}\left(\frac{\ddot\omega_{\rm ref}}{\omega_{\rm ref}}-\frac{3}{2}\left(\frac{\dot\omega_{\rm ref}}{\omega_{\rm ref}}\right)^2\right)\, ,
\end{equation}
which turns out to be
the external evolution that leads to a STA.

In the context of quantum field theory under the influence of time dependent backgrounds,  free fields can be described as a set of interacting harmonic oscillators with time-dependent frequencies. Assuming that the time dependence takes place in a finite time period, the departure from adiabaticity
is measured by the Bogoliubov transformation that connects the IN and OUT Fock spaces. When non-trivial, this transformation indicates the presence of particle creation, which is related to the friction work mentioned before. In the particular case of quantum fields in flat spacetime with time dependent boundary conditions and/or moving boundaries this effect is named motion induced radiation or DCE \cite{DCE}.

A naive generalisation of the WKB approach described above, applied to each harmonic oscillator,  does not work in general, because the effective frequency $\omega_{\rm eff}(t)$ turns out to
be ``mode-dependent", that is, it produces a STA for a given mode, but not to the full field. However, for a conformal field in $1+1$ dimensions, it is possible to follow a different approach to the STA,   based on  conformal transformations.  We will consider a massless scalar field confined to a cavity of variable size $L(t)$. We assume the 
left mirror located at $x=0$ and the right mirror at $x=L(t)$, where we impose Dirichlet boundary conditions.  The idea is to perform a coordinate conformal transformation such that, in the new coordinates, both mirrors are at rest. 
The transformation is given by \cite{Moore}
\begin{align}\label{cct}
    & \bar t +\bar x = R(t+x)\nonumber\\
    &\bar t -\bar x = R(t-x)\, ,
\end{align}
where the function $R$ is fixed in such a way that $x=0$ corresponds to $\bar x=0$ and $x=L(t)$ to
$\bar x = 1$. These conditions imply that $R$ must satisfy the so called Moore equation \cite{Moore}
\begin{equation}\label{Mooreq}
    R(t+L(t))-R(t-L(t))=2\, .
\end{equation}

Due to the conformal invariance of the classical action, the Klein-Gordon equation for the massless 
scalar field is the usual wave equation in the new coordinates. 
The field modes can be written as
\begin{eqnarray}\label{Mooremodes}
f_{n}(x,t)&=&\frac{1}{\sqrt{4\pi n}}\big(\exp\left(-i n \pi R(t+x)\right)\nonumber\\ 
&-& \exp\left(-i n \pi R(t-x)\right)\big).
\end{eqnarray}
Being functions of $t\pm x$, these modes are solutions of the wave equation. Moreover, the Moore equation implies that the modes satisfy Dirichlet boundary conditions
on the mirrors.

For a static cavity of size $L_0$, the general solution to the Moore equation reads $R(t)=t/L_0 + r(t)$,
where $r(t)$ is an arbitrary function of period $2 L_0$. The usual modes for the static cavity are
obtained setting $r(t)=0$.
Assuming that the right mirror is at rest in the IN and OUT regions, i.e. for $t\to\pm\infty$,  we can define the
IN Moore function as the solution of Eq.\eqref{Mooreq} such that $R_{\rm IN}(t)\to t/L_{0}$ as
$t\to -\infty$, and  $R_{\rm OUT}$
as the solution that satisfies
$R_{\rm OUT}(t)\to t/L_{1}$ as
$t\to +\infty$, where  $L_0$ and $L_1$ are the initial and final sizes of the cavity, respectively.
These functions determine the IN and OUT modes through Eq.\eqref{Mooremodes}. In general we will have that, as $t\to + \infty$,   $R_{\rm IN}(t)=R_{\rm OUT}(t) + r(t)$, with a non-vanishing periodic function $r(t)$. The IN and OUT field modes
will be connected by a non-trivial Bogoliubov transformation, which describes physically the creation
of particles induced by the motion of the right mirror, or DCE, and is encoded in the non-vanishing function 
$r(t)$.

To evaluate the time evolution of the mean value of the energy-momentum tensor in a thermal state, one
 can follow the traditional approach based on point-splitting regularisation.
 The result is \cite{Fulling, Birrel}
\begin{align}
\langle T_{tt}\rangle  &= \langle T_{xx}\rangle = G(t-x)+G(t+x)\nonumber\\
\langle T_{tx}\rangle  &= \langle T_{xt}\rangle = -G(t-x)+G(t+x), 
\end{align}
with
\begin{eqnarray}
G &=& -\frac{1}{24\pi}[(R'''/R')-\dfrac{3}{2}(R''/R')^2]\nonumber\\
&+& \dfrac{(R')^2}{2}[-\dfrac{\pi}{24}+F(TL_0)]    \, ,
\end{eqnarray}
where $F(TL_0)$ is related to the initial thermal energy of the cavity
\begin{equation}
F(TL_0)=\sum_{n\geq 1}\frac{n\pi}{(\exp(\frac{n\pi}{L_0 T}-1)}\, .
\end{equation}
These equations are a slight generalisation of the zero temperature results in \cite{Fulling}, see also Ref.\cite{Alves}.
For the static case, the terms proportional to $(R')^2$  in the different components of the stress-tensor reproduce the Casimir energy density and force in $1+1$ dimensions.

The total energy inside the cavity is given by
\begin{equation}
  E(t)=\int_0^{L(t)}  \langle T_{tt}(x,t)\rangle\, dx\, ,
\end{equation}
while for an adiabatic evolution the energy is obtained using the approximation $R'(t)\simeq 1/L(t)$ and neglecting derivatives of $L(t)$:
\begin{equation}
  E_{\rm ad}(L,T)=-\frac{\pi}{24 L} + \frac{F(TL_0)}{L}\, .  
\end{equation}

The above results suggest a simple way to introduce STA in our conformal field theory, using an
``inverse engineering'' approach: any Moore function $R$ that satisfies at the same time the two conditions
$R\to t/L_0 $ as $t\to -\infty$ and $R\to t/L_1 $ as $t\to + \infty$ will produce an evolution
in which the Bogoliubov transformation between the IN and OUT bases is the identity \cite{Castagnino}. Therefore, there will be no particle creation, and the occupation numbers will be the same in the initial and final states.
The associated trajectory is defined implicitly by the Moore Eq. \eqref{Mooreq}. If $R$ is defined as a piecewise function, it should be smooth enough in order to  avoid divergences in the mean value of the stress tensor.

In order to make contact with previous approaches for STA in other contexts, it is useful to discuss the adiabatic (or WKB) solutions for the field modes. For an adiabatic trajectory $L(t)$, the Moore function
is approximately given by \cite{Moore}
\begin{equation}
    R_{\rm WKB}[L]=\int^t \frac{dt'}{L(t')}\, .
\end{equation}
Let us consider a (non-adiabatic) reference trajectory $L_{\rm ref}(t)$
that starts at $L_0$ and ends at $L_1$. We would like to find
an effective trajectory, $L_{\rm eff}(t)$, such that the time evolution of the modes is the
WKB - like solution  evaluated on $L_{\rm ref}(t)$. If we insert $R_{\rm WKB}[L_{\rm ref}]$ in the field
modes Eq.\eqref{Mooremodes}, the modes will satisfy exactly the wave equation, but will not satisfy the
Dirichlet boundary conditions on $x=L_{\rm ref}(t)$.  Instead, the modes will vanish on an effective trajectory
that satisfies
\begin{equation}\label{efftraj}
\int_{0}^{t+L_{\rm eff}(t)}\frac{1}{L_{\rm ref}(t')}dt'-\int_{0}^{t-L_{\rm eff}(t)}\frac{1}{L_{\rm ref}(t')}dt'=2.
\end{equation}
We stress that this equation defines the effective trajectory $L_{\rm eff}(t)$.
Therefore, implementing the effective trajectory, the field modes will be described by the WKB modes of the reference trajectory $L_{\rm ref}$. Moreover, the initial and final lengths of the cavity will be the same for both trajectories, and the population of the modes
will also be the same at the initial and final times: $L_{\rm eff}(t)$ is the adiabatic shortcut associated with reference trajectory $L_{\rm ref}$, just as $\omega_{\rm eff}$ is the STA for the quantum harmonic oscillator with frequency $\omega_{\rm ref}$.

It is simple to see that,  if $t+L_{\rm eff}(t)<0$,  then $L_{\rm eff}(t)=L_{\rm ref}(0)=L_0,\ \text{for }t<-L_{\rm ref}(0)$.
Analogously, if $t-L_{\rm eff}(t)>\tau$ then
$L_{\rm eff}(t)=L_{\rm ref}(\tau)=L_1,\ \text{for }t>\tau+L_{\rm ref}(\tau).$
This tells us that this effective trajectory is static before $t=-L_0$
and after  $t=\tau+L_1$. Moreover, it coincides with
the reference trajectory $L_{\rm ref}$ before and after. A relevant question
is whether this is in fact a real trajectory meaning that its speed is
always below $c=1$. We can answer this question taking the time derivative
of the defining equation and solving for the speed of the wall
\begin{equation}
\frac{d}{dt}L_{\rm eff}(t)=\frac{L(t+L_{\rm eff}(t))-L(t-L_{\rm eff}(t))}{L(t-L_{\rm eff}(t))+L(t+L_{\rm eff}(t))}\leq1,
\end{equation}
from which we see that its speed is indeed bounded by the speed of
light.

\begin{figure}
	\begin{center}
		\includegraphics[scale=0.53]{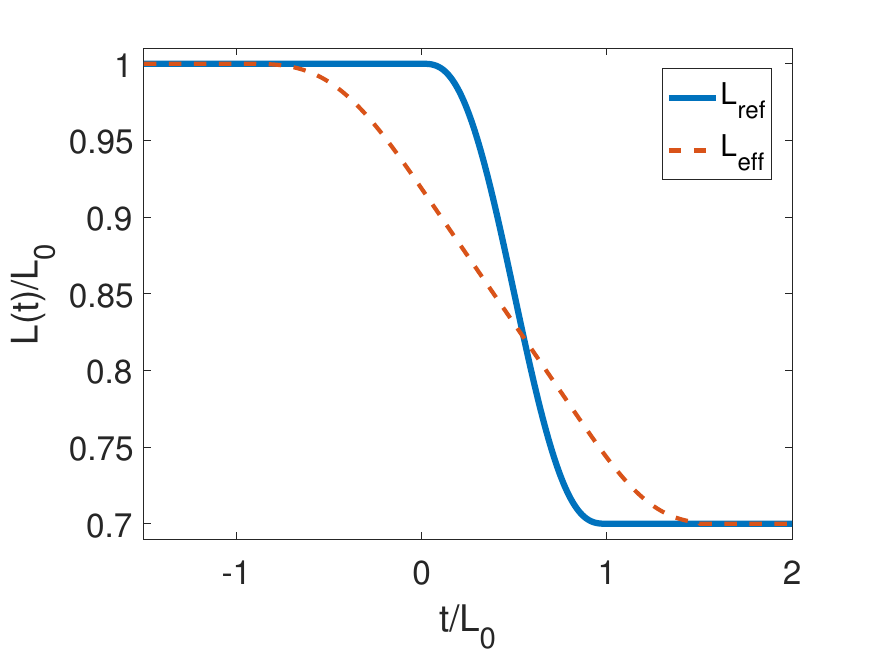}
		\caption{ A reference trajectory (blue solid line) given by Eq. (\ref{eq:L}) with $\epsilon=0.3$ and $\tau/L_0=1$. The corresponding effective trajectory (orange dashed line) calculated from Eq. (\ref{efftraj}) gives the shortcut to adiabaticity.}
		\label{fig:leff}
	\end{center}
\end{figure}

\begin{figure}
	\begin{center}
		\includegraphics[scale=0.35]{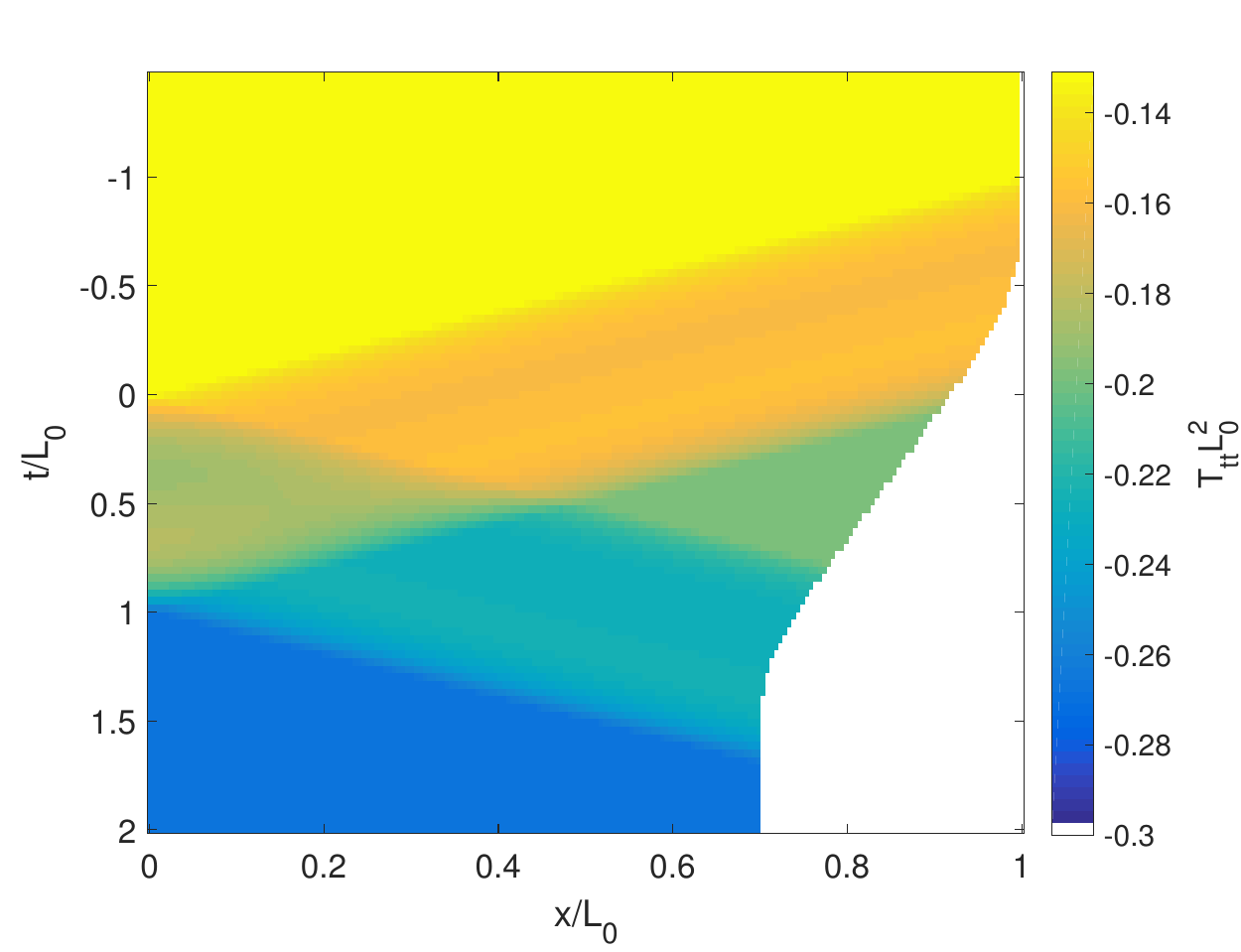}
		\caption{ Energy density for an adiabatic shortcut corresponding to the polynomial trajectory with length $L_1/L_0=0.7$ and $\tau/L_0=1$ from $t=-L_0$ to $t=L_1+\tau$ and the field initially in the vacuum state. The energy density is a negative constant before and after the compression and it is smaller at the end.}
		\label{fig:Edensityatajo}
	\end{center}
\end{figure}

\begin{figure}
		\includegraphics[width=0.8\linewidth]{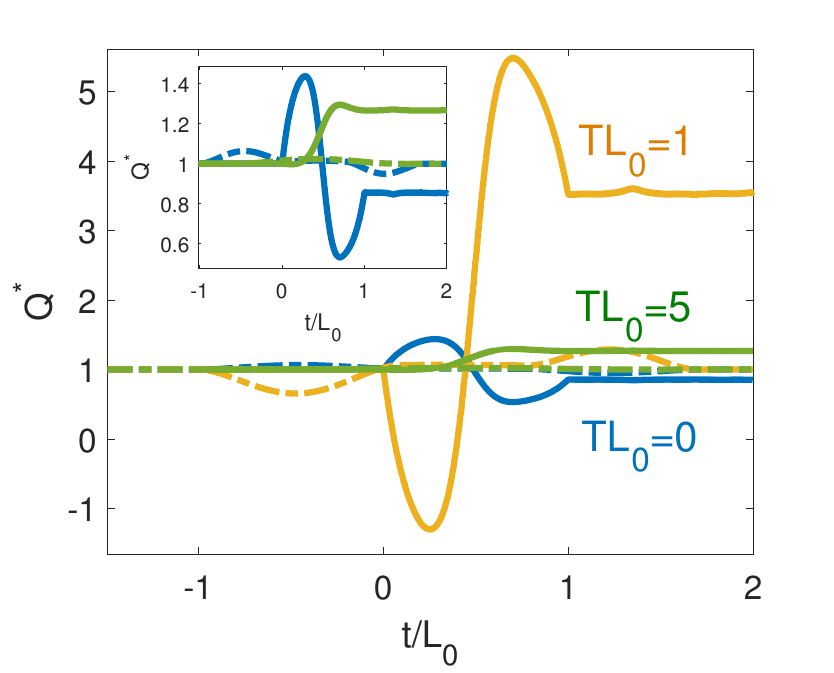}%
	\caption{Adiabaticity parameter as a function of time for the reference (solid line) and effective trajectory (dashed line) at three different temperatures. An inset plot has been added to better visualize the curves for $TL_0=0$ and $TL_0=5$. The parameters used for the reference trajectory were $\epsilon=0.3$ and $\tau/L_0=1$.}
	\label{fig:Q}
\end{figure}

A simple smooth trajectory that interpolates between $L_0$ and $L_1$ is given by 
\begin{equation}
\label{eq:L}
    L_{\rm ref}(t)=\begin{cases}
\begin{array}{c}
L_{0}\, ,\\
L_0\, (1-\epsilon \delta(t))\, ,\\
L_{1}\, ,
\end{array} & \begin{array}{c}
t<0\\
0<t<\tau\\
\tau<t \, ,
\end{array}\end{cases}.
\end{equation}
where $\epsilon$ is a positive constant less than 1, the final distance is given by $L_1=L_0(1-\epsilon)$ and the evolution function is
\begin{align}
\label{eq:poli}
\delta(t)=10(t/\tau)^3-15(t/\tau)^4+6(t/\tau)^5.
\end{align}
In Fig.\ref{fig:leff} we show the effective trajectory $L_{\rm eff}$ associated to this $L_{\rm ref}$. The energy density inside the cavity evolves from the static Casimir thermal energy corresponding to a cavity of length $L_0$ to that corresponding to $L_1$. The evolution is non-adiabatic at intermediate times, as depicted in Fig.\ref{fig:Edensityatajo} for the particular case $T=0$: indeed, for an adiabatic evolution the energy density would be constant inside the cavity at each time, and equal to the static Casimir energy density corresponding to the instantaneous size of the cavity.   Note also that, for an arbitrary trajectory (not a shortcut), the final energy density would also contain the contribution of the created particles.  In Fig. \ref{fig:Q}, 
we plot the adiabaticity parameter $Q^*(t)=E(t)/E_{\text{ad}}(t)$ for the cavity as a function of time, during the shortcut,  at different temperatures for the reference trajectory as well as for the corresponding STA. This parameter effectively measures the distance between the adiabatic energy for an infinitely slow motion and the actual energy in the cavity for the STA found, being of course equal to 1 when these two coincide. We can see that although both of them depart from the adiabatic result at intermediate times, the STA returns to 1 at the end of the motion confirming the implementation of an adiabatic shortcut and its deviation from adiabaticity at intermediate times is relatively small. Although there are certain times at which the reference trajectories cross $Q^*=1$ this does not constitute a STA, since a sudden stop of the moving wall at those times would generate DCE photons by itself, that would make $Q^*\neq1$ afterwards. This is why it is crucial that the trajectories considered have a null final velocity.

To summarise, solving the equation for $L_{\rm eff}(t)$ we get a trajectory
that, when implemented,  generates an evolution of the state of the quantum
field that in the end coincides with that of an exact adiabatic evolution
along $L_{\rm ref}(t)$. That is,  we have found a shortcut to adiabaticity
for a scalar quantum field in a cavity with a moving wall. The results can be  generalized to other two dimensional conformal fields confined in a cavity of variable length, as a massless Dirac field satisfying bag boundary conditions on the boundaries \cite{MPC87}.

\section{Implications on the quantum Otto cycle}

In order to discuss some fundamental limits for the power and efficiency
of a quantum  Otto cycle, 
we will consider the  following limiting trajectory $L_{\rm ref}(t)=L_0\theta(-t)+L_1\theta(t)$, which corresponds to a very small displacement time, $\tau\ll L_0,L_1$.

In this case we can replace in Eq. (\ref{efftraj}) to find that the effective trajectory is given by 
\begin{equation}
    L_{\rm eff}(t)=\begin{cases}
\begin{array}{c}
L_{0}\\
\frac{2L_{0}L_{1}-t(L_{0}-L_{1})}{L_{0}+L_{1}}\\
L_{1}
\end{array} & \begin{array}{c}
t<-L_{0}\\
-L_{0}<t<L_{1}\\
L_{1}<t
\end{array}\end{cases}.
\end{equation}
This means in the limit case where we want to implement an instantaneous length change without spurious photon generation, we need at least $L_0+L_1$ units of real time to implement it and it consists of a linear motion. This is consistent with previous results for this particular case \cite{Fulling, Castagnino}. 

\begin{figure}
	\begin{center}
		\includegraphics[scale=0.5,trim={3cm 8cm 4cm 8cm},clip]{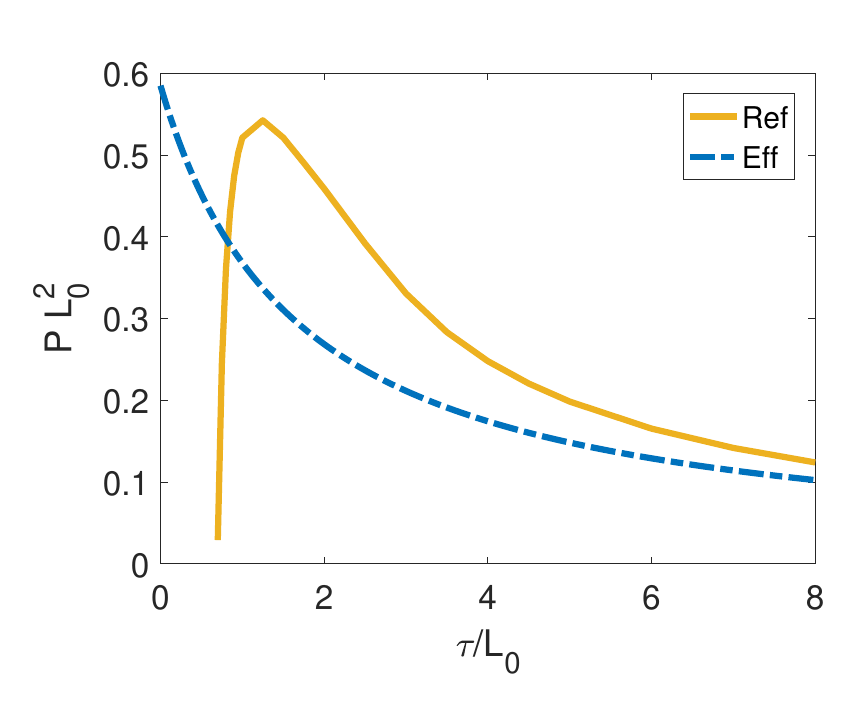}
		\caption{ Power for an Otto cycle as a function of the timescale, $\tau$, implementing the reference (solid line) or effective STA trajectory (dashed line) for the expansion and compression strokes. The parameter used  for the trajectory was $\epsilon=0.3$, while the thermal baths used for the cycle had temperatures $T_0L_0=1$ and $T_1L_0=5$.}
		\label{fig:P}
	\end{center}
\end{figure}

We have previously mentioned that finite time driving is necessary to improve the power delivered but usually results in additional friction energy on the working medium that diminishes the efficiency of the engine. In the case of a scalar quantum field with a moving boundary we have found a STA, which  maximises the work delivered, that turns out to be $\langle w_{\text{ad}}\rangle$,  and the efficiency, $\eta=\eta_{\text{ad}}$. Moreover,  the power produced is bounded by the minimum time that it takes to implement the STA (twice for compression and expansion and under the assumption that the thermalisation times are much shorter than the compression and expansion ones) \begin{equation}
P=\frac{ W_{\text{ad}}}{2(L_0+L_1+\tau)}\leq\frac{ W_{\text{ad}}}{2(L_0+L_1)}.
\end{equation}
In Fig. \ref{fig:P} we compare the power given by an engine under a quantum Otto cycle whose expansion and compression strokes given either by the reference or the effective STA  trajectory. We can see that for slow motions (ie, large $\tau$) they converge to the same value, since $\tau\gg L_0$ and $W\approx W_{\text{ad}}$, but for extremely fast motion the power of the reference trajectory decreases rapidly and becomes negative while the STA, which always has a superior efficiency, also provides a higher power. Perturbative calculations to second order in $\epsilon$ also show that the power associated with the reference trajectory decays as $1/\tau^4$ for short cycle times \cite{Otto_nos} in concordance with the figure presented here.

Even though we have given an explicit simple way to compute an STA that maximises both the efficiency and the power delivered, it may be very difficult to implement in practice. For example for an Otto cycle, the efficiency is given by
\begin{equation}
    \eta_{\text{otto}}=1-\frac{L_1}{L_0}\leq\eta_{\text{carnot}}=1-\frac{T_1}{T_0},
\end{equation}
and is bounded by Carnot's efficiency. Then given a cavity of length $L_0$ and thermal baths $T_1$ and $T_0$, if we want to achieve the best possible efficiency the final length is fixed to $L_1/L_0=T_1/T_0$. Implementing then the best possible STA forces us to move the mirror at the speed $v=(L_0-L_1)/(L_0+L_1)$, which for $T_0\gg T_1$ gives $L_0\gg L_1$,  and the speed $v\approx 1$ approximates the speed of light, leading to an experimental impossibility.

Alternatively, we can set the problem with the constraint that the speed of the mirror should be at most a constant $v$. Then, to get the most possible power we would take an STA with constant speed $v=(L_0-L_1)/(L_0+L_1)$. This would set the final length to $L_1=L_{0}(1-v)/(1+v)$. We can now clearly see that the efficiency 
\begin{equation}
    \eta_{\text{otto}}=1-\frac{(1-v)}{(1+v)}
\end{equation}
is directly bounded by the maximum speed the mirror can reach.

We now address some related problems:   Is it possible to perform a translation of the cavity
avoiding the excitation of the modes? For a non-rigid cavity this can be trivially achieved 
by a two step STA, moving first the right mirror and then the left mirror, both following STA. This translation without excitation may be relevant for discussions on relativistic quantum information where it has been shown the DCE can degrade the entanglement between observers in relative motion \cite{RQI}. More generally, the STA protocol presented here could be useful for other conformal field theories satisfying time dependent boundary conditions, providing a route for modifying these conditions without altering the state of the field in a bounded region.

\section{Discussion}

Is it possible to have a STA for a single accelerated mirror? In other words, what happens with the population of the field modes outside the cavity? In this case there are no STA: assuming that the mirror is initially static,  accelerates during some time, and 
finally becomes at rest, 
the total energy outside the cavity is strictly positive \cite{Fulling}, and  given by the energy of the created particles \cite{Walker}. Unlike
what happens inside the cavity, the DCE cannot be reversed.

Finally, we describe a possible experimental realisation of the shortcut. Superconducting circuits have proved to be useful to simulate a one dimensional cavity with a moving mirror \cite{Nori}. Indeed, the DCE has been observed experimentally using  a superconducting waveguide ended by a SQUID, and the time dependent external condition is implemented by varying the magnetic flux $\Phi(t)$ on the SQUID \cite{Delsing}. In order to perform a STA, one should consider a closed waveguide ended by a SQUID, and apply an effective time dependent 
magnetic flux $\Phi_{\rm eff}(t)$. When the field is initially in the vacuum state, no photons should be detected after applying the magnetic field on the SQUID. Indeed, we have recently proposed to implement an Otto cycle in this system \cite{Otto_nos}, and showed that, for expansion and compression of small amplitude, it is possible to avoid the DCE and maximise the efficiency of the cycle. The results of this paper show that the DCE can be avoided for trajectories of arbitrary amplitude. Although there is no net generation of photons that would decrease the efficiency, there is instantaneous emission and reabsortion at intermediate times. This represents an energy cost that should be further investigated in the future, since previous works \cite{cost1,third1,third2,Deffner3} have associated this cost with the third law of thermodynamics and quantum speed limits. \\

\section*{Acknowledgements}

This research was supported by Agencia Nacional de Promoción Científica y Tecnológica (ANPCyT), Consejo Nacional de Investigaciones Científicas y Técnicas (CON- ICET), Universidad de Buenos Aires (UBA) and Univer- sidad Nacional de Cuyo (UNCuyo). P.I.V. acknowledges ICTP-Trieste Associate Program.

\end{document}